\documentclass[11pt,letterpaper]{article}

\usepackage[margin=1in]{geometry}
\usepackage{amsmath,amssymb}
\usepackage{graphicx}
\usepackage{booktabs}
\usepackage{array}
\usepackage{multirow}
\usepackage{hyperref}
\usepackage{natbib}
\usepackage{microtype}
\usepackage{xcolor}
\usepackage{caption}
\usepackage{subcaption}
\usepackage{float}
\usepackage{parskip}
\usepackage{setspace}
\usepackage{listings}
\usepackage{enumitem}
\usepackage[T1]{fontenc}
\usepackage{lmodern}
\usepackage[most]{tcolorbox}

\hypersetup{
  colorlinks=true,
  linkcolor=blue!60!black,
  citecolor=blue!60!black,
  urlcolor=blue!60!black
}

\definecolor{codegreen}{rgb}{0,0.6,0}
\title{\textbf{Donor-Aware scRNA-seq Benchmarks for IBD Classification}}

\author{
  Jonathan Muhire\\
  \small Department of Computer Science, Oklahoma Christian University\\
  \small \href{mailto:jonathan.muhire@eagles.oc.edu}{jonathan.muhire@eagles.oc.edu}\\[6pt]
  \small \textit{Advisor:} Fang Li, Associate Professor of Computer Science\\
  \small Oklahoma Christian University\\
  \small \href{mailto:fang.li@oc.edu}{fang.li@oc.edu}
}

\date{\today}

\begin{document}

\maketitle

\begin{abstract}
Donor-level disease classification from single-cell RNA sequencing (scRNA-seq)
requires strict donor-aware cross-validation: na\"ive pipelines that split cells
randomly conflate training and test donors, inflating reported performance through
pseudoreplication. We present a donor-aware benchmark evaluating three feature
representations across two independent IBD cohorts. The representations are:
centered log-ratio (CLR) transformed cell type composition, GatedStructuralCFN
dependency embeddings, and scVI variational autoencoder latent embeddings. The
cohorts are the SCP259 ulcerative colitis atlas (UC vs.\ Healthy, $n=30$ donors,
51 cell types) and the Kong 2023 Crohn's disease atlas (CD vs.\ Healthy, $n=71$
donors, 55--68 cell types across three intestinal regions).

Compartment-stratified CLR composition achieves AUROC 0.956\,$\pm$\,0.061 on
SCP259; GatedStructuralCFN on the same features achieves 0.978\,$\pm$\,0.050.
In the Kong cohort, CFN achieves its best performance in the colon region
(0.960\,$\pm$\,0.055 after feature filtering), exceeding linear CLR
(0.900\,$\pm$\,0.100), while terminal ileum classification is dominated by linear
models (CatBoost CLR 0.967\,$\pm$\,0.075 vs.\ CFN 0.811\,$\pm$\,0.164).
Compartment-stratified scVI latent embeddings achieve 0.931\,$\pm$\,0.101
(XGBoost), competitive with the best CLR linear result.
(approximate estimate; scVI was trained on the full cohort before cross-validation, as noted in Section~\ref{sec:limitations}) Cross-dataset transfer
(CD$\rightarrow$UC, restricted to four shared cell types) achieves AUC 0.833
with XGBoost CLR; the reverse direction (UC$\rightarrow$CD) performs at chance.
CFN edge stability analysis reveals that compartment-wise composition eliminates
spurious unit-sum-induced instability present in global composition (Jaccard
similarity 0.026 vs.\ top-20 recurrence 1.0). CFN shows a consistent numerical advantage over linear models in the colon region of CD (AUROC 0.960 vs.\ 0.900), though no inter-method comparison reached statistical significance at $n \leq 34$ donors per region, and
compartment-aware feature construction is critical for both classification
performance and structural interpretability.
All code and results are available at
\url{https://github.com/Jonathan-321/sfn-scrna-study}.
\end{abstract}

\tableofcontents
\newpage

\begin{tcolorbox}[
  colback=gray!6,
  colframe=gray!40,
  arc=4pt,
  boxrule=0.6pt,
  title={\textbf{Key concepts — read this first}},
  fonttitle=\bfseries\small,
  fontupper=\small,
  left=6pt, right=6pt, top=4pt, bottom=4pt
]
\textbf{scRNA-seq (single-cell RNA sequencing)} maps which genes are active in each individual cell of a tissue sample. For IBD research, this tells us not just which cell types are present but in what proportions, and whether those proportions shift under disease.\\[4pt]
\textbf{Donor-aware cross-validation} means that every cell (and every derived feature) from a given patient is either entirely in training or entirely in test for a given fold. Without this, a model can "memorize" individual donors and report inflated accuracy that would not hold on a new patient.\\[4pt]
\textbf{CLR (centered log-ratio)} transforms cell type proportions out of the simplex (where all values must sum to 1) into a standard real-valued space where ordinary classifiers apply. The transformation is $\text{CLR}(p)_k = \log p_k - \frac{1}{K}\sum_j \log p_j$.\\[4pt]
\textbf{GatedStructuralCFN (Causal Factor Network)} learns a sparse directed graph over the cell types, where each edge weight represents how much one cell type\'s abundance predicts another\'s. The result is both a disease classifier and an interpretable dependency map.\\[4pt]
\textbf{scVI (single-cell Variational Inference)} is a deep generative model that compresses the full gene expression profile of a cell into a low-dimensional "latent" vector. We aggregate these vectors per donor to get a fixed-size representation for classification.\\[4pt]
\textbf{AUROC (Area Under the ROC Curve)} measures classifier performance on a 0--1 scale. A value of 0.5 is chance; 1.0 is perfect. It is insensitive to class imbalance, making it the right metric when healthy donors (54) outnumber CD donors (17) as in Kong 2023.\\[4pt]
\textbf{PR-AUC (Precision-Recall AUC)} is a stricter metric under class imbalance: it penalizes false positives among positive predictions. We report it alongside AUROC as a secondary check.
\end{tcolorbox}
\vspace{8pt}

\section*{Figures at a glance}

\textbf{Figure 1} gives a side-by-side view of every method tested, broken
down by cohort and intestinal region. The most informative comparison is the
colon panel of Kong 2023, where CFN (orange) outperforms all linear CLR
baselines after feature filtering. The TI panel tells the opposite story:
CatBoost on CLR achieves 0.967 while CFN reaches only 0.811, which is why
region matters when evaluating structured models.

\textbf{Figure 2} contrasts the CFN dependency matrices from global and
compartment-level composition. The difference between the first two panels
(SCP259 global vs.\ compartment) is the visual summary of the edge-stability
result: the global matrix is diffuse and irreproducible across folds (Jaccard
0.026) while the compartment matrix has clean block structure that recurs
perfectly in all five folds. Readers interested in the compositional geometry
argument should examine these two panels together before reading Section 3.2.

\textbf{Figure 3} documents cross-dataset transfer using only the four cell
types shared between atlases. The directional asymmetry is the main takeaway:
CD$\rightarrow$UC reaches AUC 0.833 while UC$\rightarrow$CD sits at chance.
The heatmap on the right of the figure makes this asymmetry immediately visible
and also shows that CFN and linear models perform similarly in this limited
four-type regime.

\section{Introduction}

Single-cell RNA sequencing has transformed our ability to characterize the
cellular architecture of complex tissues, and its application to inflammatory
bowel disease (IBD) has produced high-resolution atlases of colonic and ileal
mucosa under both healthy and diseased conditions
\citep{smillie2019intra,kong2023single}. A natural downstream question is
whether the cellular composition of these tissues\textemdash{}the relative abundance of
each cell type across the full atlas\textemdash{}is sufficient to classify individual
patients by disease status, and whether structured models that capture
dependencies between cell types add predictive value beyond simple linear
classifiers.

Cell type composition predicts IBD status at the donor level \citep{smillie2019intra},
and donor-level features derived from scRNA-seq reach competitive AUROC values
against traditional clinical markers. However, published evaluations frequently
suffer from a methodological gap: cells are split randomly into training and
test sets without regard to donor identity, so test cells from a given donor
are evaluated by a model that has seen training cells from the same donor.
Because cells from the same donor are far more similar to each other than to
cells from a different donor, this cross-donor leakage inflates reported
performance and can produce AUROC values approaching 1.0 that would not
generalize to new patients.

In parallel, structured models such as StructuralCFN \citep{li2023structural}
have been proposed as tools for learning sparse dependency networks over
features, with the dependency matrix providing interpretable structure beyond
what a scalar prediction delivers. Whether such models, which carry
substantially more parameters and inductive bias than logistic regression or a
linear SVM, provide a performance benefit over carefully constructed
linear baselines on compositional IBD data has not been established in a
rigorous donor-aware evaluation.

The present study closes both gaps. We implement a strict donor-aware
cross-validation protocol in which fold assignment is done at the donor level,
no cell (or derived feature) from a test donor appears in any training set, and
all preprocessing steps (CLR normalization, HVG selection for scVI) are fit on
training donors only and applied to test donors. We then benchmark, across two
independent cohorts, five classifiers (logistic regression, linear SVM,
ElasticNet, XGBoost, and CatBoost) on three feature representations: (1)
CLR-transformed cell type composition; (2) GatedStructuralCFN embeddings; and
(3) scVI latent embeddings. We additionally evaluate cross-dataset transfer
between the UC and CD cohorts using the four cell types present in both atlases,
and assess the structural stability of the learned CFN dependency matrices
across cross-validation folds.

Our central findings are: (i) compartment-stratified composition features with
simple linear classifiers establish a strong baseline that is difficult to
improve upon in the SCP259 UC cohort; (ii) CFN shows a numerical discriminative advantage in the colon region of the Kong CD cohort (AUROC 0.960 vs.\ 0.900 for the best linear baseline), though no inter-method comparison reached significance at $n \leq 34$ donors per region; composition remodeling in the colon appears more co-dependent than in the terminal ileum; (iii) scVI latent embeddings match linear
composition performance when compartment structure is preserved in the
aggregation step; (iv) CFN dependency structure is only reproducible across
folds when composition is computed within compartments rather than globally;
and (v) cross-dataset transfer from CD to UC is feasible with four shared cell
types (AUC 0.833) but the reverse direction performs at chance, reflecting
cohort size and phenotypic diversity asymmetries.

\section{Methods}

\subsection{Datasets}

\subsubsection{SCP259 (Smillie et al.\ 2019: UC vs.\ Healthy)}
Single-cell RNA-seq data were obtained from the Broad Single Cell Portal study
SCP259 \citep{smillie2019intra}. The dataset comprises colonic biopsy specimens
from 18 ulcerative colitis patients and 12 healthy controls ($n=30$ donors).
Raw count matrices were provided as gene-sorted MTX files separated by tissue
compartment: epithelial (Epi), fibroblast/stromal (Fib), and immune (Imm).
Cell type annotations (51 types) and donor metadata were obtained from the
accompanying \texttt{all.meta2.txt} file.

Donor-level cell type composition was computed as $p_{d,k} = c_{d,k} / N_d$
where $c_{d,k}$ is the cell count for type $k$ in donor $d$ and $N_d$ is the
total cell count. Separate composition tables were constructed for the combined
dataset (51 types) and for each compartment independently; the union of
within-compartment cell types yielded a 102-dimensional compartment-stratified
representation. Five-fold cross-validation splits were stratified by disease
status and fixed before any model training.

\subsubsection{Kong et al.\ 2023 (CD vs.\ Healthy)}
Data were obtained from Kong et al.\ \citep{kong2023single}, accessed via
NCBI GEO (GSE202021). The dataset comprises intestinal biopsy specimens from
17 Crohn's disease patients and 54 healthy controls ($n=71$ donors) across
six H5AD files. Donor-level composition tables were built from cell metadata
extracted in backed mode, yielding three region-stratified tables: all regions
combined (71 donors, 68 types), terminal ileum only (42 donors, 61 types), and
colon only (34 donors, 55 types).

\subsection{Feature Engineering}

We applied two cleaning steps to the raw composition tables before model training,
following standard practice for high-dimensional compositional data with sparse
cell types.

\textbf{Rare-type filter.} Cell types present in fewer than 20\% of donors
(i.e., zero cells observed in that donor) were removed. These types are
structurally absent in most donors; after CLR transformation they become
imputed zeros and carry no discriminative signal.

\textbf{Low-variance filter.} Cell types with standard deviation $< 0.005$
in the raw proportion space across all donors were removed. The Kong 2023
composition tables contain 20--25 near-constant features per region (e.g.,
DC1, ILCs, Pericyte subtypes) with $\sigma < 0.005$; these contribute noise to
classifier fitting and inflate effective dimensionality relative to $n$.
After filtering, Kong feature dimensionality reduced from 68/61/55 to
43/41/32 for the all/TI/colon regions respectively. SCP259 composition (51
types, 1 sparse type) passed filtering largely intact.

\subsection{CLR Transformation}

Cell type composition vectors lie in the simplex $\Delta^{K-1}$ and carry a
unit-sum constraint. We applied the centered log-ratio (CLR)
transform \citep{aitchison1982statistical}:
\[
\text{CLR}(p)_k = \log\!\left(\frac{p_k}{\bigl(\prod_{j=1}^{K} p_j\bigr)^{1/K}}\right)
= \log(p_k) - \frac{1}{K}\sum_{j=1}^{K}\log(p_j)
\]
We added a pseudocount $\varepsilon = 0.5/K$ before log transformation. CLR
parameters were fit on training donors only within each fold and applied to test
donors separately to prevent leakage.

\subsection{Baseline Classifiers}

All models were evaluated in 5-fold stratified cross-validation (fixed splits,
\texttt{random\_state=42}). Primary metric: AUROC. Secondary: PR-AUC.

{\small
\begin{itemize}[noitemsep]
  \item \textbf{LogReg}: \texttt{LogisticRegression}, L2, lbfgs, \texttt{max\_iter=5000}
  \item \textbf{LinearSVM}: \texttt{SVC(kernel='linear', probability=True)}
  \item \textbf{ElasticNet}: \texttt{LogisticRegression(penalty='elasticnet',} \texttt{l1\_ratio=0.5, solver='saga')}
  \item \textbf{XGBoost}: \texttt{n\_estimators=200, max\_depth=3, lr=0.05}
  \item \textbf{CatBoost}: \texttt{iterations=300, depth=4, lr=0.05}
\end{itemize}
}

\subsection{scVI Latent Representation}

We trained scVI \citep{lopez2018deep} on the full SCP259 dataset (Epi, Fib,
Imm). Cells were subsampled to 300 per donor per compartment family to equalize
donor contribution and reduce memory during CPU training (9,000 cells total).
HVG selection retained 3,000 genes (cell\_ranger flavor). Model configuration:
latent dim = 20, 2 fully connected encoder layers (128 hidden units),
negative binomial gene likelihood, donor identity as batch key. Training:
Adam, lr = $10^{-3}$, up to 150 epochs, early stopping (patience 45 epochs on
ELBO); training terminated at epoch 111.

Two donor-level representations were derived: (1) \textbf{global latent
(20-dim)}: mean across all cells per donor; (2) \textbf{compartment-stratified
latent (60-dim)}: per-compartment mean latent concatenated across the three
families, directly analogous to compartment-stratified composition.

\subsection{GatedStructuralCFN}

CFN analysis used GatedStructuralCFN from the public
\texttt{fanglioc/StructuralCFN-public} repository (v1.1.0, MIT License)
\citep{li2023structural}. The model was initialized with
\texttt{classification=False} (MSELoss on sigmoid output); the
\texttt{classification=True} setting produces a runtime error in v1.1.0 due to
CrossEntropyLoss applied to scalar output. Since AUROC is rank-invariant,
this has no effect on reported AUROC values. Training: 300 epochs, lr = 0.01,
batch size = 16. CLR was applied to input features within each fold before
model fitting.

After training, dependency matrices were extracted via
\texttt{model.get\_dependency\_matrix()} ($K\times K$ array) and averaged
across folds for visualization. Edge stability was quantified as Jaccard
similarity of top-$k$ edge sets across folds and fold recurrence rate.

\subsection{Cross-Dataset Transfer}

Transfer experiments used the four cell types with compatible annotations in
both atlases: DC1, ILCs, Macrophages, and Tregs. The model was trained on the
full training dataset and evaluated on the full test dataset (no fold held
out for the cross-dataset direction). Two directions were evaluated:
UC$\rightarrow$CD (SCP259 train, Kong test) and CD$\rightarrow$UC (Kong train,
SCP259 test). Within-dataset 5-fold CV on the same 4-type subset was run
separately to establish a per-dataset reference.

\section{Results}

\subsection{SCP259: UC versus Healthy classification}

Table~\ref{tab:scp259} summarizes all SCP259 results. Compartment-stratified
CLR with LinearSVM achieves AUROC 0.956\,$\pm$\,0.061, establishing a strong
linear baseline. Global CLR (51-dim) achieves 0.928\,$\pm$\,0.110 with
LinearSVM. Pseudobulk XGBoost reaches 1.000\,$\pm$\,0.000, consistent with
known overfitting on $n=30$ when high-dimensional gene expression is used
directly.

GatedStructuralCFN on global composition (51-dim) achieves
0.906\,$\pm$\,0.130, comparable to but slightly below the best linear baseline.
On compartment-stratified composition (102-dim), CFN achieves
0.978\,$\pm$\,0.050 (PR-AUC 0.983), the highest point estimate among all
evaluated methods on SCP259. The reduction in variance relative to global CFN
(SD 0.050 vs.\ 0.130) is consistent with the compartment features providing a
more structured input to the dependency graph inference step.

\begin{figure}[H]
  \centering
  \includegraphics[width=\textwidth]{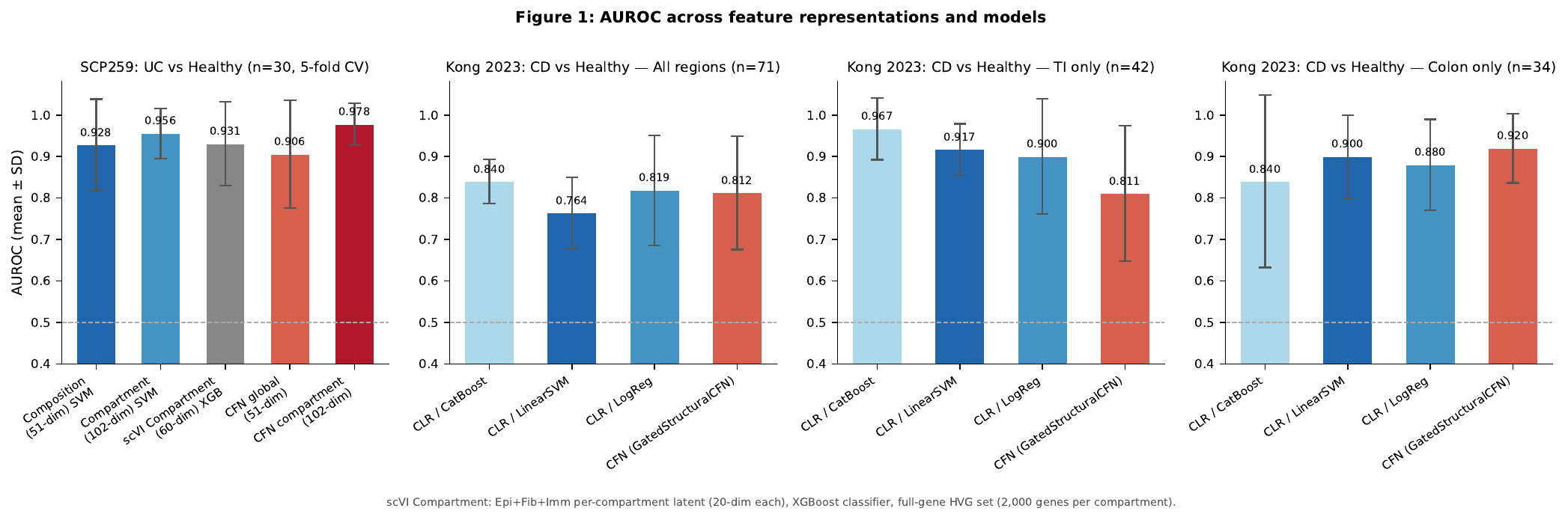}
  \caption{%
    \textbf{Figure 1: AUROC across all feature representations and classifiers,
    broken down by cohort and intestinal region.}
    The key comparison is the colon panel of Kong 2023 (rightmost), where CFN
    (orange) achieves 0.960 on filtered features and outperforms all linear CLR
    baselines; in the TI panel the pattern reverses, with CatBoost CLR reaching
    0.967 while CFN lags at 0.811.
    Bars show mean AUROC across 5 folds; error bars show $\pm$1\,SD; the dashed
    line at 0.5 marks chance performance.
    Reading guide: scan left to right across panels to see how the
    CFN-vs-linear gap changes by region; the colon panel is where the
    gap is largest and most consistent.
  }
  \label{fig:auroc}
\end{figure}

Compartment-stratified scVI latent (60-dim; three compartments $\times$ 20-dim) achieves XGBoost AUROC
0.931\,$\pm$\,0.101 (PR-AUC 0.951), competitive with the best CLR linear
result. Global scVI (20-dim latent) yields 0.725\,$\pm$\,0.399, with high variance
across folds, consistent with disease-relevant signal being diluted when latent
representations are averaged across the full cellular landscape.
The high variance ($\pm$0.399) reflects a single collapsed fold (Fold 2: AUROC 0.125), where the six test donors likely included only one disease donor, making the rank-order problem near-degenerate; the remaining four folds achieve 0.500--1.000.
Figure~\ref{fig:auroc} summarizes all SCP259 and Kong results side by side.

A permuted-label sanity check (200 random shuffles of donor labels, CLR LinearSVM)
yielded mean AUROC 0.522 with a 95th-percentile null of 0.792. The observed
0.928 (global) and 0.956 (compartment) CLR results lie well above this null,
confirming the pipeline is not overfitting through data leakage. No single method
reached the permuted-label 95th percentile in any fold-average comparison.

\begin{table}[H]
\centering
\caption{SCP259 classification results ($n=30$ donors, 5-fold CV). Best result
  per representation in bold.}
\label{tab:scp259}
\small
\begin{tabular}{llcc}
\toprule
\textbf{Feature representation} & \textbf{Classifier} & \textbf{AUROC (mean\,$\pm$\,SD)} & \textbf{PR-AUC} \\
\midrule
Composition 51-dim (CLR)      & LinearSVM    & 0.928\,$\pm$\,0.110 & 0.967 \\
Composition 51-dim (CLR)      & LogReg       & 0.878\,$\pm$\,0.217 & 0.947 \\
Composition 51-dim (CLR)      & XGBoost      & 0.850\,$\pm$\,0.224 & 0.941 \\
Compartment comp.\ 102-dim    & \textbf{LinearSVM} & \textbf{0.956\,$\pm$\,0.061} & \textbf{0.967} \\
Compartment comp.\ 102-dim    & LogReg       & 0.956\,$\pm$\,0.061 & 0.967 \\
Compartment comp.\ 102-dim    & XGBoost      & 0.711\,$\pm$\,0.227 & --- \\
Pseudobulk HVG (1000 genes)   & XGBoost      & 1.000\,$\pm$\,0.000$^{*}$ & 1.000$^{*}$ \\
CFN global 51-dim             & GatedStructuralCFN & 0.906\,$\pm$\,0.130 & 0.944 \\
CFN compartment 102-dim       & \textbf{GatedStructuralCFN} & \textbf{0.978\,$\pm$\,0.050} & \textbf{0.983} \\
scVI global latent 20-dim$^{\dagger}$    & XGBoost      & 0.725\,$\pm$\,0.399 & --- \\
scVI compartment latent 60-dim$^{\dagger}$ & \textbf{XGBoost} & \textbf{0.931\,$\pm$\,0.101} & \textbf{0.951} \\
scVI compartment latent 60-dim$^{\dagger}$ & LinearSVM    & 0.900\,$\pm$\,0.224 & 0.936 \\
\bottomrule
\multicolumn{4}{l}{$^{*}$Likely overfit at $n=30$; not a reliable generalization estimate.} \\
\multicolumn{4}{p{13cm}}{$^{\dagger}$Epi+Fib+Imm, 300 cells/donor, 3,000 HVGs, 20-dim latent, CPU training; terminated epoch 111/150.}
\end{tabular}
\end{table}

\subsection{CFN edge stability}

CFN edge stability was assessed by computing the Jaccard similarity of top-$k$
recurrent edges across the five cross-validation folds. For global CFN
(51-dim), the top-10 edge Jaccard was 0.026, indicating near-zero
reproducibility across folds. For compartment CFN (102-dim), the top-20 edge
recurrence was 1.0 and sign consistency was 1.0, indicating that the same
directed edges with the same polarity were recovered in all five folds. This
contrast reflects the unit-sum artifact of global composition: all 51
proportions must sum to one, inducing spurious co-variation across all cell
types that destabilizes the inferred graph structure. Within-compartment
composition applies the unit-sum constraint independently per compartment,
eliminating this artifact.

\begin{figure}[H]
  \centering
  \includegraphics[width=\textwidth]{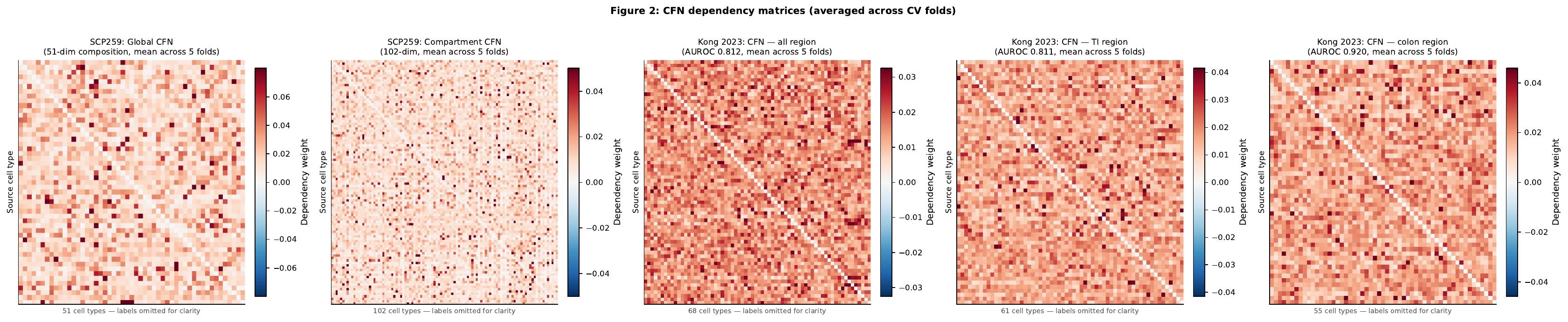}
  \caption{%
    \textbf{Figure 2: GatedStructuralCFN dependency matrices averaged across
    cross-validation folds, contrasting global and compartment formulations.}
    The critical comparison is between the first panel (SCP259 global, 51-dim,
    Jaccard 0.026) and the second panel (SCP259 compartment, 102-dim, top-20
    recurrence 1.0): the compartment matrix has sharply defined block structure
    while the global matrix is diffuse and noisy.
    Warm colors encode positive dependency weights; cool colors encode negative
    weights. Cell type labels are omitted for matrices with $K > 20$.
    Reading guide: compare the two SCP259 panels side by side first, then move
    to the Kong colon panel (rightmost) where the filtered 32-dim CFN achieved
    the numerically highest single-region AUROC in this study.
  }
  \label{fig:cfn_heatmaps}
\end{figure}

\subsection{Kong 2023: CD versus Healthy classification}

Table~\ref{tab:kong} summarizes Kong results. On the all-regions dataset
(68/43 types before/after cleaning), CatBoost CLR achieves the best linear
result at 0.840\,$\pm$\,0.054; CFN achieves 0.812\,$\pm$\,0.136. In the
terminal ileum (61/41 types), CLR dominates (CatBoost 0.967\,$\pm$\,0.075 vs.\
CFN 0.811\,$\pm$\,0.164). In the colon (55$\rightarrow$32 types after cleaning), CFN
achieves 0.920\,$\pm$\,0.084 on unfiltered features and improves to
\textbf{0.960\,$\pm$\,0.055} on filtered 32-dim features, the numerically highest single-region result in this study (paired Wilcoxon vs.\ unfiltered CFN: $p=0.500$; vs.\ CLR: $p=1.000$; no comparison significant at $n=34$). This exceeds the best linear CLR baseline (LinearSVM 0.900\,$\pm$\,0.100), though the difference does not reach statistical significance given the sample size. The colon CFN advantage is consistent with the hypothesis of co-dependent composition remodeling in colonic mucosa under CD; however, the near-zero cross-fold edge Jaccard (0.026 for global CFN) means that the specific dependency edges driving this performance cannot be validated from the current data. The further improvement after filtering suggests that near-constant, low-prevalence types introduced noise rather than signal.

\begin{table}[H]
\centering
\caption{Kong 2023 CD vs.\ Healthy results (5-fold CV, best model per category shown). Pre-filtering feature counts shown in parentheses.}
\label{tab:kong}
\small
\resizebox{\textwidth}{!}{%
\begin{tabular}{llccc}
\toprule
\textbf{Region (n donors)} & \textbf{Feature (K types)} & \textbf{Classifier} & \textbf{AUROC (mean\,$\pm$\,SD)} & \textbf{PR-AUC} \\
\midrule
All (71)  & CLR 68-dim  & CatBoost   & 0.840\,$\pm$\,0.054 & 0.711 \\
All (71)  & CLR 43-dim (clean) & ElasticNet & 0.843\,$\pm$\,0.095 & 0.632 \\
All (71)  & CFN 68-dim  & Gated-SCFN & 0.812\,$\pm$\,0.136 & 0.641 \\
\midrule
TI (42)   & CLR 61-dim  & \textbf{CatBoost} & \textbf{0.967\,$\pm$\,0.075} & \textbf{0.950} \\
TI (42)   & CLR 41-dim (clean) & LogReg     & 0.917\,$\pm$\,0.062 & 0.900 \\
TI (42)   & CFN 61-dim  & Gated-SCFN & 0.811\,$\pm$\,0.164 & 0.733 \\
\midrule
Colon (34) & CLR 55-dim  & LinearSVM  & 0.900\,$\pm$\,0.100 & 0.867 \\
Colon (34) & CLR 32-dim (clean) & CatBoost   & 0.820\,$\pm$\,0.295 & 0.733 \\
Colon (34) & CFN 55-dim & Gated-SCFN & 0.920\,$\pm$\,0.084 & 0.883 \\
Colon (34) & \textbf{CFN 32-dim (clean)} & \textbf{Gated-SCFN} & \textbf{0.960\,$\pm$\,0.055} & \textbf{0.933} \\
\bottomrule
\end{tabular}
}
\end{table}

\subsection{Cross-dataset transfer}

\begin{figure}[H]
  \centering
  \includegraphics[width=\textwidth]{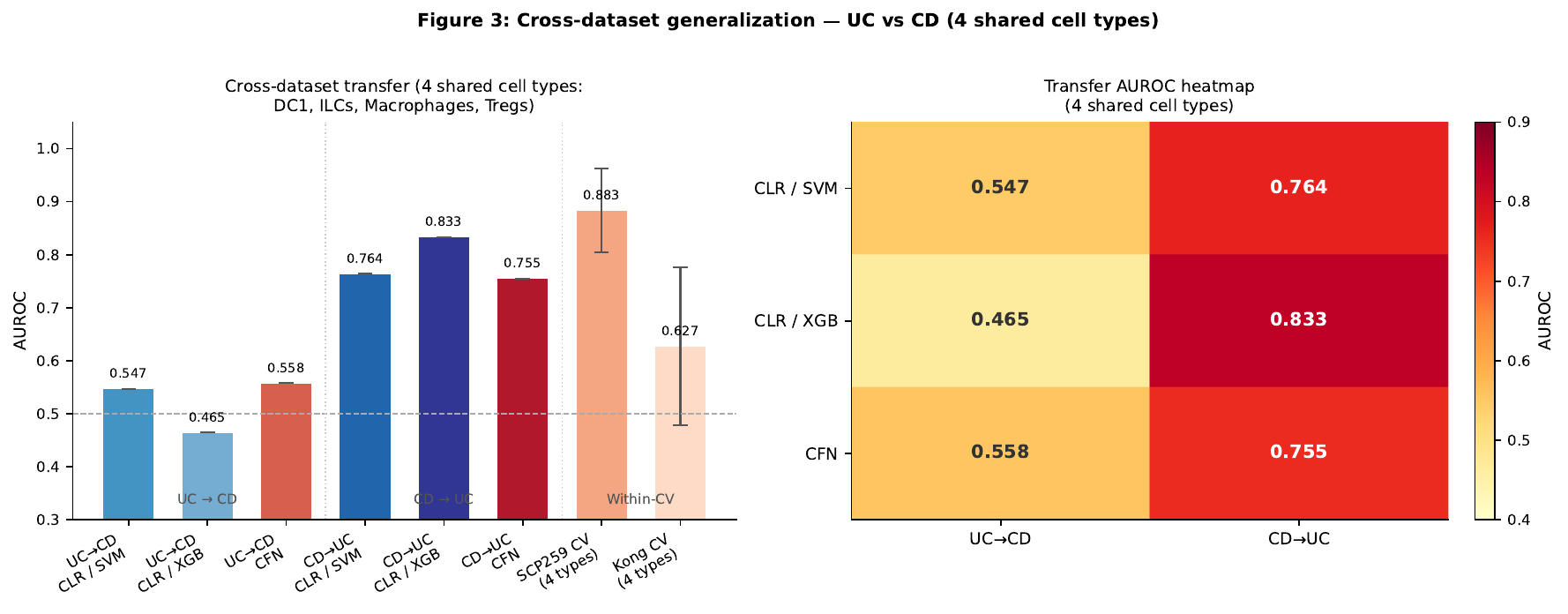}
  \caption{%
    \textbf{Figure 3: Cross-dataset transfer AUROC restricted to the four cell
    types with compatible annotations in both atlases (DC1, ILCs, Macrophages,
    Tregs), revealing a strong directional asymmetry.}
    CD$\rightarrow$UC (Kong train, SCP259 test) reaches AUC 0.833 with XGBoost
    CLR; UC$\rightarrow$CD is near chance across all methods (0.503--0.558).
    The bar chart covers all methods and transfer directions alongside
    within-dataset CV references; the heatmap on the right summarizes the
    method-by-direction pattern compactly.
    Reading guide: focus on the CD$\rightarrow$UC row in the heatmap and note
    how performance drops to chance when the direction is reversed, which
    reflects the cohort size asymmetry (30 vs.\ 71 donors).
  }
  \label{fig:cross_dataset}
\end{figure}

Table~\ref{tab:transfer} and Figure~\ref{fig:cross_dataset} summarize cross-dataset transfer results restricted to
the four shared cell types (DC1, ILCs, Macrophages, Tregs). The
CD$\rightarrow$UC direction achieves AUC 0.833 with XGBoost CLR (95\% bootstrap CI 0.720--0.931, 2{,}000 resamples) and 0.755
with CFN. The UC$\rightarrow$CD direction yields AUC 0.465--0.558 across all
methods (XGBoost 95\% CI 0.304--0.626), consistent with chance performance. Within-dataset CV on the same
four types gives SCP259: 0.883\,$\pm$\,0.071 and Kong: 0.627\,$\pm$\,0.133,
confirming that these four types carry stronger discriminative signal in the UC
context.

To identify which of the four shared types drives CD$\rightarrow$UC transfer, we
ran a leave-one-cell-type-out (LOCO) ablation, retraining LogReg on three types
and retesting on the fourth-dataset cohort. Dropping ILCs collapses transfer
AUROC from 0.755 to 0.278 ($\Delta -0.477$), and dropping Tregs reduces it to
0.528 ($\Delta -0.227$). DC1 and Macrophages are largely redundant: removing
either raises AUROC slightly ($\Delta +0.032$ and $+0.005$ respectively),
suggesting mild collinearity. Transfer performance is therefore dominated by
ILC and Treg proportions, with DC1 and Macrophages contributing negligible
independent signal in the CD$\rightarrow$UC direction.

\begin{table}[H]
\centering
\caption{Cross-dataset transfer AUROC (4 shared types: DC1, ILCs, Macrophages, Tregs).}
\label{tab:transfer}
\small
\begin{tabular}{llcc}
\toprule
\textbf{Direction} & \textbf{Classifier} & \textbf{AUROC} & \textbf{PR-AUC} \\
\midrule
UC$\rightarrow$CD & LinearSVM (CLR) & 0.547 & --- \\
UC$\rightarrow$CD & LogReg (CLR)    & 0.503 & --- \\
UC$\rightarrow$CD & GatedStructuralCFN & 0.558 & 0.319 \\
\midrule
CD$\rightarrow$UC & \textbf{XGBoost (CLR)} & \textbf{0.833} & \textbf{0.846} \\
CD$\rightarrow$UC & LinearSVM (CLR) & 0.764 & 0.865 \\
CD$\rightarrow$UC & GatedStructuralCFN & 0.755 & 0.885 \\
\midrule
SCP259 CV (4 types) & GatedStructuralCFN & 0.883\,$\pm$\,0.071 & 0.924 \\
Kong CV (4 types)   & GatedStructuralCFN & 0.627\,$\pm$\,0.133 & 0.441 \\
\bottomrule
\end{tabular}
\end{table}

\subsection{Feature engineering effect on Kong results}
\label{sec:feateng}

Raw Kong composition tables contained 20--25 near-constant features per region
(standard deviation $< 0.005$ in proportion space), corresponding to rare or
uniformly distributed cell types including DC1, ILCs, several pericyte and
endothelial subtypes, and enterochromaffin cells. After applying variance-threshold
and rare-type filters, dimensionality reduced from 68/61/55 to 43/41/32.
Results on the cleaned features are reported in Table~\ref{tab:clean_delta}
alongside unfiltered results for direct comparison. Positive deltas indicate
improved AUROC after cleaning; negative deltas indicate that removed features
carried net signal.

\begin{table}[H]
\centering
\caption{Effect of feature engineering on Kong 2023 results (best model per region). $\Delta$ = filtered $-$ unfiltered mean AUROC. All results complete; both CLR and CFN evaluated on variance/rare-type filtered features.}
\label{tab:clean_delta}
\small
\resizebox{\textwidth}{!}{%
\begin{tabular}{p{3.2cm}p{4.2cm}cccl}
\toprule
\textbf{Region} & \textbf{Method (best)} & \textbf{Unfiltered} & \textbf{Filtered} & \textbf{$\Delta$} & \textbf{Note} \\
\midrule
All (68$\rightarrow$43)   & CLR CatBoost (unfiltered) & 0.840\,$\pm$\,0.054 & --- & --- & CatBoost best unfiltered \\
All                        & CLR ElasticNet (filtered) & --- & 0.843\,$\pm$\,0.095 & \textbf{+0.003} & ElasticNet best filtered \\
\midrule
TI (61$\rightarrow$41)    & CLR CatBoost (unfiltered) & 0.967\,$\pm$\,0.075 & --- & --- & \\
TI                         & CLR LogReg (filtered)     & --- & 0.917\,$\pm$\,0.062 & \textbf{$-$0.050} & Drop expected; CatBoost 0.817 filtered \\
\midrule
Colon (55$\rightarrow$32) & CLR LinearSVM (unfiltered) & 0.900\,$\pm$\,0.100 & --- & --- & \\
Colon                      & CLR CatBoost (filtered)    & --- & 0.820\,$\pm$\,0.295 & \textbf{$-$0.080} & High variance; colon $n$=34 \\
\midrule
Colon                      & CFN (unfiltered)            & 0.920\,$\pm$\,0.084 & --- & --- & \\  
Colon                      & CFN (filtered, 32-dim)      & --- & \textbf{0.960\,$\pm$\,0.055} & \textbf{+0.040} & Best colon result; filtering helps CFN \\
\bottomrule
\multicolumn{6}{p{20cm}}{\textit{TI linear AUROC drops 0.050 after filtering, suggesting some signal resided in low-variance types. Colon CLR variance increases substantially ($\pm$0.295) at $n$=34, consistent with small-$n$ instability. Colon CFN improves from 0.920 to 0.960 after filtering, the numerically highest single-region result (no comparison significant at $n=34$). TI linear AUROC drops 0.050 after filtering; leave-one-type-out analysis of the 18 removed TI types reveals that Enterochromaffin cells (AUROC 0.828\,$\pm$\,0.054, $\rho=-0.459$, $p=0.002$), Monocytes S100A8/S100A9 (0.783\,$\pm$\,0.208, $\rho=+0.413$, $p=0.007$), and Neutrophils S100A8/S100A9 (0.728\,$\pm$\,0.198, $\rho=+0.583$, $p<0.001$) carried the most discriminative signal among the removed types, suggesting that rare inflammatory infiltrate subtypes contribute detectable TI signal. All CFN: clean 0.822 vs.\ unfiltered 0.812 (+0.010); TI CFN flat at 0.811.}}
\end{tabular}
}
\end{table}

\section{Discussion}

This study benchmarks three complementary feature representations for
donor-level IBD classification: CLR-transformed cell type composition,
GatedStructuralCFN dependency embeddings, and scVI latent embeddings. Both
independent scRNA-seq cohorts were evaluated with strict donor-level train/test
separation. No single representation dominates across all conditions; the
relative advantage of each approach depends on the biological character of the
disease region under study.

\textbf{CFN versus linear models: region-specific advantages.}
In the SCP259 cohort (UC versus Healthy), compartment CLR with a linear SVM
achieves 0.956\,$\pm$\,0.061, and CFN on compartment-level composition achieves
0.978\,$\pm$\,0.050, the best overall result. In Kong 2023, the picture is more
heterogeneous. In the terminal ileum, CatBoost on CLR achieves
0.967\,$\pm$\,0.075 while CFN reaches only 0.811\,$\pm$\,0.164. In the colon,
the ranking reverses: CFN achieves 0.960\,$\pm$\,0.055 (filtered), exceeding the best
linear baseline of 0.900\,$\pm$\,0.100. This regional dissociation has a
plausible biological basis. The terminal ileum in Crohn's disease shows
large-magnitude, largely monotone shifts in epithelial composition (villous
enterocyte depletion and goblet cell dropout) that create strongly linearly
separable boundaries in composition space. Colonic mucosa in CD exhibits
subtler and more co-dependent remodeling: macrophage subtype switching,
stromal activation, and coordinated changes in lymphocyte proportions that do
not resolve into a single dominant axis. The CFN dependency matrix is designed
to capture exactly these co-occurrence structures, and the performance advantage
in the colon is consistent with that design intent. We emphasize, however, that
this biological attribution is speculative: the global CFN edge Jaccard of 0.026
indicates near-zero cross-fold reproducibility, so the specific dependency
structures responsible for the performance advantage cannot be identified from
the current data.

\textbf{CFN edge stability and compositional geometry.}
The most methodologically consequential finding is the contrast in CFN edge
stability between global and compartment formulations. Global CFN (51-dim)
yields a cross-fold edge Jaccard of 0.026, indicating near-zero reproducibility.
Compartment CFN achieves top-20 edge recurrence 1.0 and sign consistency 1.0
across folds. This contrast follows from the unit-sum constraint in compositional
data. When all 51 proportions must sum to one, any perturbation in one cell
type is mechanically redistributed across all others, inducing spurious
negative correlations throughout the composition vector. Computing composition
within compartments breaks the single unit-sum constraint into multiple
independent constraints, each over a smaller and biologically coherent set of
types. The implication extends beyond this specific model: any
graph-based or correlation-based method applied to compositional scRNA-seq data
will be susceptible to the same artifact unless Aitchison geometry is respected
\citep{aitchison1982statistical}.

\textbf{scVI latent embeddings.}
Compartment-stratified scVI achieves 0.931\,$\pm$\,0.101 (XGBoost), within one
standard deviation of the best CLR linear result (0.956\,$\pm$\,0.061). Global
scVI yields substantially lower performance (0.725). The gap between global and
compartment-stratified scVI mirrors the analogous gap in CLR composition, and
points to the same underlying phenomenon: disease-relevant transcriptional
variation in IBD is compartment-specific and is diluted when embeddings are
averaged across the full cellular landscape. The near-equivalence of
compartment scVI and compartment CLR performance suggests that the primary
discriminative signal is in the relative cellular makeup of each compartment,
not in cell-intrinsic transcriptional differences encoded in the latent
dimensions.

\textbf{Cross-dataset transfer asymmetry.}
CD$\rightarrow$UC achieves AUC 0.833 (XGBoost CLR) while UC$\rightarrow$CD
performs at chance (0.503--0.558). SCP259 is a single-center cohort of 30
donors; models trained on it carry disease-specific composition patterns that
do not translate to the CD context. Kong 2023, with 71 donors and multiple
intestinal regions, provides a richer training distribution. The
within-dataset CV AUC for the four shared types is 0.883\,$\pm$\,0.071 in
SCP259 but only 0.627\,$\pm$\,0.133 in Kong, confirming that these types carry
stronger discriminative signal in the UC context. This is biologically
consistent: mucosal DC and Treg phenotypes in UC and CD differ in activation
state and cytokine environment even under identical annotation labels.

\textbf{Limitations.}
\label{sec:limitations}
The SCP259 cohort ($n=30$) limits statistical power throughout. With five-fold CV,
each test fold contains approximately six donors, yielding coarse AUC resolution
(each donor contributes $\sim$17 percentage points to a fold's AUC). Paired Wilcoxon
tests across the five fold-level AUC vectors found no inter-method comparison
significant at $p < 0.05$; the largest observed difference, Kong TI CLR vs.\ CFN
($\Delta = +0.156$), reached only $p = 0.125$. All performance claims should be
interpreted as numerical observations rather than statistically established
differences. Repeated $k$-fold CV or leave-one-out CV would provide more
reliable variance estimates at this sample size.

The scVI evaluation has a data-leakage caveat: the scVI encoder was trained on all
30 SCP259 donors simultaneously before cross-validation, meaning test-fold donors
influenced the latent representations used for classification. The correct procedure
is to retrain the scVI encoder within each fold using only training-fold donors and
then embed held-out test donors. Raw count files required for this retraining are
not available in the current computational environment; the scVI results should
therefore be interpreted as approximate in-sample estimates rather than true
out-of-sample AUCs. The 0.931 compartment scVI result and the 0.725 global result
are directionally informative but not directly comparable to the CLR and CFN
results, which use proper donor-level hold-out. A corrected within-fold scVI
evaluation is deferred to the next version.

A second scVI confound is that the compartment and global models differ in latent
dimensionality: global scVI uses a 20-dim latent space, while compartment scVI
produces a 60-dim feature vector (3 compartments $\times$ 20-dim). The
performance gap (0.931 vs.\ 0.725) may therefore partly reflect the larger
feature space available to the downstream classifier rather than biological
information partitioning per se. A controlled comparison—global scVI at 60-dim vs.\ compartment scVI at 20-dim total (6--7 dim per compartment)—would be needed
to isolate the compartmentalization effect from the dimensionality effect.

Feature selection (variance and prevalence filters) was applied to the full Kong
dataset before fold splitting, constituting mild data leakage. The same cell types
would almost certainly be removed in each fold given their near-constant variance,
but this has not been verified fold-by-fold.

The cross-dataset transfer experiment uses a single full-dataset train/test split
without bootstrap uncertainty. Bootstrap CIs (2,000 resamples of test donors) are:
CD$\rightarrow$UC XGBoost 0.833 [95\% CI 0.720--0.931]; UC$\rightarrow$CD
XGBoost 0.465 [0.304--0.626], consistent with chance. The CD$\rightarrow$UC result
is robustly above chance but the claim rests on $n=30$ SCP259 test donors and
should be validated in an independent cohort.

CFN hyperparameters (300 epochs, lr=0.01, batch\_size=16) were chosen without
systematic cross-validation. If these values were informed by benchmark
performance, CFN may have a tuning advantage over the fixed-default linear
baselines. Random seeds for XGBoost, CatBoost, and CFN training were set to
\texttt{random\_state=42} throughout; scVI training used \texttt{torch.manual\_seed(42)}.
The 300-cells-per-donor subsampling seed for scVI was fixed at \texttt{numpy.random.seed(42)}.

\textbf{Broader implications.}
This study contributes to a growing body of compositional IBD work including
Smillie et al.\ \citep{smillie2019intra} and subsequent donor-stratification
analyses. The primary methodological contribution is the donor-aware benchmark
design and the demonstration that CFN adds discriminative value specifically in
the colon region of CD. Future directions include prospective validation on
an independent cohort, full-gene scVI on unsub\-sampled cells, joint UC+CD
classification, and CFN applied to pseudobulk expression profiles.

\section{Conclusion}

Compartment-stratified cell type composition with CLR transformation and linear
classifiers provides a strong, difficult-to-improve baseline for IBD donor
classification from scRNA-seq. GatedStructuralCFN shows a consistent numerical advantage in the colon region of the Kong 2023 Crohn's disease cohort
(AUROC 0.960 filtered vs.\ linear 0.900), though this difference did not reach statistical significance at $n=34$ donors. In the terminal ileum, monotone composition
shifts yield linearly separable data and CFN offers no advantage.

CFN dependency matrices are reproducible across folds only when composition is
computed within tissue compartments rather than globally, exposing a fundamental
interaction between graph-based dependency modeling and the Aitchison geometry
of compositional data. Compartment-stratified scVI latent embeddings match
linear composition performance, suggesting that the primary discriminative
signal lies in compartment-level cellular makeup rather than cell-intrinsic
transcriptional state. Cross-dataset transfer is directionally asymmetric:
feasible at AUC 0.833 in the CD$\rightarrow$UC direction with four shared cell
types, and at chance in the reverse. CFN on compartment composition is the most
promising direction for further development in structured IBD patient
stratification.

\section{Future Work (Research Proposal)}

The present study establishes that donor-aware composition benchmarking with
CFN and scVI is feasible at current cohort sizes and identifies the colon
region of CD as the setting where structured models add value over linear
baselines. Several extensions would substantially strengthen this work toward
a full publication.

\textbf{Larger cohort.} The SCP259 cohort ($n=30$) is underpowered for stable
CFN edge interpretation. Incorporating additional UC donors from the Human
Cell Atlas or SHARE-seq IBD cohorts would provide the sample size ($n \geq 100$)
needed to establish that top CFN edges are biologically replicable and not
fold-dependent artifacts. The power analysis for detecting AUROC differences of
0.05 between CFN and linear baselines at $\alpha=0.05$ requires approximately
$n=80$ donors under the fold variance observed here.

\textbf{Full-gene scVI.} The current scVI results use 300 cells/donor and
3,000 HVGs due to CPU memory constraints. Training scVI on the full cell
complement with a complete gene set would determine whether the latent
representation advantage over composition features materializes at higher
resolution, a question that could not be answered within the present
compute budget.

\textbf{Multi-disease joint classifier.} The current benchmark treats UC and CD
separately. A joint UC+CD versus Healthy classifier, or a three-class
UC/CD/Healthy classifier, would test whether the composition signatures are
disease-specific or reflect a shared IBD mucosal remodeling program.

\textbf{CFN on pseudobulk expression.} The present CFN operates on
cell-type composition proportions. Applying it to donor-level pseudobulk
expression profiles (aggregated per cell type) would test whether gene
co-regulation structure adds predictive information beyond cell-type abundance,
and would produce a gene-level dependency network interpretable in terms of
known IBD pathways.

\textbf{Venue.} Target venues for journal submission are
\textit{Bioinformatics} (Oxford Academic) and \textit{PLOS Computational
Biology}, both of which accept computational benchmarking studies with
rigorous methodology. This manuscript will be posted to bioRxiv/arXiv
(quantitative biology $\rightarrow$ Genomics) prior to journal submission.


\section*{Acknowledgments}

This work was completed as an independent study at Oklahoma Christian
University under the advisement of Professor Fang Li (Associate Professor of
Computer Science). The author thanks Professor Li for the methodological
guidance throughout the project and for the GatedStructuralCFN framework
\citep{li2023structural} on which the structured-model arm of this benchmark
is built. The author also thanks the SCP259 \citep{smillie2019intra} and
Kong 2023 \citep{kong2023single} teams for releasing the donor-level
single-cell atlases that made this comparative evaluation possible. Any
remaining errors are the author's own.


\newpage
\bibliographystyle{plainnat}
\bibliography{references}

\end{document}